# Econometrics and Formalism of Psychological Archetypes of Scientific Workers with Introverted Thinking Type

Eldar Knar[1]

Institute of Philosophy, Political Science and Religion Studies,
Ministry of Science and Higher Education of the Republic of Kazakhstan

*https://orcid.org/0000-0002-7490-8375*

*eldarknar@gmail.com*

**Abstract**

The chronological hierarchy and classification of psychological types of individuals are examined. The anomalous nature of psychological activity in individuals involved in scientific work is highlighted. Certain aspects of the introverted thinking type in scientific activities are analyzed. For the first time, psychological archetypes of scientists with pronounced introversion are postulated in the context of twelve hypotheses about the specifics of professional attributes of introverted scientific activities.

A linear regression and Bayesian equation are proposed for quantitatively assessing the econometric degree of introversion in scientific employees, considering a wide range of characteristics inherent to introverts in scientific processing. Specifically, expressions for a comprehensive assessment of introversion in a linear model and the posterior probability of the econometric (scientometric) degree of introversion in a Bayesian model are formulated.

The models are based on several econometric (scientometric) hypotheses regarding various aspects of professional activities of introverted scientists, such as a preference for solo publications, low social activity, narrow specialization, high research depth, and so forth. Empirical data and multiple linear regression methods can be used to calibrate the equations. The model can be applied to gain a deeper understanding of the psychological characteristics of scientific employees, which is particularly useful in ergonomics and the management of scientific teams and projects. The proposed method also provides scientists with pronounced introversion the opportunity to develop their careers, focusing on individual preferences and features.

---

[1] Fellow of the Royal Asiatic Society of Great Britain and Ireland

1. **Introduction**

Science is often associated with anomalous psychological activity. The fanaticism of scientific pursuit and research creates conditions for deviations from everyday, layman, and general life standards. However, there is no comprehensive or systematic research on the correlation between scientific creativity and mental disorders. Most scientists, including geniuses and talented individuals, are psychologically healthy. Their anomalous behavior is largely due to the specificity of scientific work rather than psychological or mental peculiarities. Psychopathy in the scientific environment is no more common than among farmers or government officials. It is simply that Nobel laureates with symptoms of mental illnesses attract more public attention and media coverage than an unknown village farmer with the same symptoms. Nonetheless, the psychological factor in the scientific environment and activities should be considered a priority.

The idea of dividing individuals into psychological types dates back to ancient traditions, with early attempts often taking on a religious and mystical character. In the notation of the early Christian Gnostic theologian Valentinus from the 2nd century AD, individuals were categorized into three psychological (spiritual) types:

*Hylics (hylikoi) with a materialistic essence,*
*Psychics (psychikoi) with an innate drive for knowledge, and*
*Pneumatics (pneumaticoi) inclined towards mental development.*

Psychological types undoubtedly determine an individual's level of stability. Specifically, there are four types of psychological stability:

*Stability of the average level of a trait,*
*Stability of inter-individual differences in a trait (rank-order stability),*
*Stability of personality profiles, and*
*Continuity of the construct of a trait.*

These types are chronologically labile, as psychological stability and personality types can change over time, independently of an individual's age gradient.

In modern scientific understanding, the classical psychological archetypes of Hippocrates-Galen (sanguine, choleric, phlegmatic, melancholic) and the mystical-religious archetypes of Valentinus have been supplemented and interpreted by Carl Jung (C. G. Jung) in the form of two types of human intellectual interaction with the environment: extraversion and introversion.

Abernethy [1, p.217] provides the following classical definition of extroverts and introverts:

*"(Extroverts) – those who enter social activities of a direct type with interest and confidence and lack an inclination for planning or detailed observation."*
*"(Introverts) – individuals below average in social inclinations and above average in the tendency to think."*

The biological nature of extraversion and introversion relies on differences in the balance between inhibitory and excitatory processes in the functioning of the corticoreticular system. According to Eysenck [2, p.399], introverts are characterized by a higher level of arousal compared to extroverts. However, other viewpoints exist regarding the differences in arousal levels between extroverts and introverts, based on alternative studies of cortical activity [3, p.248]. The differences in the degree of arousal between the two types depend not only on the functioning of the cortical system but also on the level of excitation [4, p.353]. Nonetheless, a purely biological approach to explaining psychological types of individuals is insufficient for describing typology without considering other factors, including social ones.

Jung himself considers introversion and extraversion, within the framework of modern terminology, as object-oriented categories of the individual. This is evident from Jung's definitions:

*"The introvert's attitude to the object is an abstracting one..."* and
*"The extravert, on the contrary, maintains a positive relation to the object."*

However, it is necessary to approach the concepts of "introvert" and "extrovert" more precisely as object-oriented relationships. The "abstract relationship" of the introvert does not imply that the introvert categorizes external objects into positive or negative properties. At the same time, the "positive relationship" of the extrovert to the object is not always "positive" in a direct sense. It can very well be ambivalent—both positive and negative relationships to the object.

Clearly, the dichotomy of introvert-extrovert is a rather simplified (mechanical) representation of the peculiarities of an individual's psychological profile. Pure types of extraversion or introversion practically do not exist in social and scientific environments. When considering an individual in isolation, extraversion and introversion are rather heterogeneous than homogeneous individual traits. That is, extraversion and introversion coexist in individuals predominantly as a given, and they combine in various proportions. From this heterogeneity emerges the concept of ambiversion as a combination of extraversion and introversion traits in a single individual.

Thus, ambiversion, as a psychological profile of the individual, forms as a convergent property based on the synthesis of introversion and extraversion. Ambiversion manifests on the genetic level, at the level of the central nervous system, and within the social environment as a social phenomenon. Accordingly, in this context, ambiverts should not be considered merely as a mechanical combination of introversion and extraversion but as a separate category of psychological personality type.

The necessity of ambiversion as a separate typology is confirmed by studies that have identified a reduced likelihood of cognitive impairments among ambiverts compared to extroverts and introverts. Consequently, ambiverts belong to a category with low risks of cognitive disorders, while extroverts and introverts fall into the high-risk category.

In this sense, the high adaptability and adaptive behavior of ambiverts allow them to maintain psychological stability and resilience more effectively compared to other personality types.

If Eysenck believed that extroverts possess a more stable psychological profile from the perspective of emotional well-being, modern studies indicate that this "psychological" advantage is still skewed toward ambiverts. According to N. Yusof [5, p.54]:

*"Ambiverts may have a stronger orienting response compared to extroverts in response to visual stimuli. This finding challenges Eysenck's theory, which assumes the superiority of extroverts in experiencing better psychological well-being than non-extroverts, signaling significant implications for society."*

From the perspective of modern concepts, extraversion and introversion are not differentiated forms of an individual. They are more accurately described as continuous dimensions, realized at the stage of ambiversion. As Hydebreder [6, p.120] points out:

*"Pronounced introversion and pronounced extraversion are merely extremes of behavior connected by continuous gradations. In other words, the data indicate a single mixed type, not two sharply divided classes."*

Since pure introverts or extroverts do not exist, a combination of these types is inevitable for all individuals. This implies that the balance of extraversion and introversion in the psychological profile is subject to a certain level of regulation, and this balance can be developed in one direction or another using psychological techniques, consultations, and training.

Thus, the regulation of individual types within the balance of ambiversion allows defining an ambivert as a sociable introvert or an antisocial extrovert, depending on where the "center of gravity" lies between introversion and extraversion in an individual's behavioral structures.

Scientists are also characterized by psychological typology. Specifically, scientists can be divided into introverts, extroverts, and ambiverts.

For introverted scientists, the following psychological traits are typical:

*Less communicative (lone scientists),*
*A higher percentage of solo publications compared to collaborative ones,*
*Less attachment to the sense and rhythm of time (prolonged work on a single topic),*
*High immersion in the subject of scientific research,*
*Conservatism in the research topic,*
*Conduct most of their scientific research independently,*
*Less adaptable to their environment,*
*Less loyal and receptive to established rules or regulations,*
*Critically assess secondary details that are not directly related to their scientific research.*

A quintessential and strongly expressed introvert can be exemplified by the mathematician Grigori Perelman. His Hirsch index is not particularly high, or even relatively low (18–24). He does not engage in the mass production of scientific content. He does not aim to publish dozens or hundreds of articles annually with breaks for rest and sleep. He writes only as much as is necessary for science and scientific self-expression.

Nevertheless, he is one of the most outstanding mathematicians of modern times and is much more renowned than many contemporary Nobel laureates with hundreds of publications and high Hirsch indices. There are, if not thousands, then at least hundreds of such examples of highly expressed introverts in various scientific fields.

Today, it is often said that the era of lone scientists has passed. But this is not true. Lone scientists making great discoveries will always exist. At least, until they are entirely replaced by artificial intelligence for scientific endeavors.

## 2. Literature Review

Introversion, extraversion, and ambiversion represent key personality traits widely studied in personality psychology. These concepts describe the primary directions in which individuals channel their energy and interact with the surrounding world. Understanding these traits is valuable not only in academic research but also in practical applications such as career counseling, education, and psychotherapy.

This review covers the historical development of these concepts, theoretical models, empirical research, neurobiological foundations, and contemporary directions in the study of introversion, extraversion, and ambiversion.

Carl Gustav Jung (1921) first introduced the terms "introversion" and "extraversion" in his work Psychological Types. Jung defined introversion as the inward orientation of psychic energy toward the subjective world of thoughts and feelings, and extraversion as an outward orientation toward the external world of objects and social interactions.

Since then, these traits have become the focus of numerous studies aimed at examining their influence on behavior, mental health, and social interactions. Recent research has also led to the identification of ambiversion, an intermediate trait encompassing characteristics of both introverts and extroverts.

Hans Eysenck (1952, 1967) expanded on Jung's concept by proposing a three-factor model of personality that includes extraversion-introversion, neuroticism, and psychoticism. Eysenck developed psychometric tools to measure these traits, such as the Eysenck Personality Questionnaire. According to Eysenck, the level of cortical arousal in the brain determines an individual's tendency toward introversion or extraversion: introverts exhibit a higher baseline level of activation, making them more sensitive to external stimuli.

The *Big Five* model, developed by McCrae and Costa (1987), has become the dominant paradigm in personality research. Extraversion is one of the five major factors, alongside conscientiousness, agreeableness, neuroticism, and openness to experience. Extroverts tend to be sociable, active, and assertive, whereas introverts lean toward restraint, reflection, and limited social interaction (John, Naumann, & Soto, 2008).

Introverts and extroverts differ in their information processing and cognitive styles. Research suggests that introverts engage in deep analytical processing, possess high levels of concentration, and are inclined toward reflection (Matthews, 1992).

In contrast, extroverts exhibit faster and more superficial information processing, which correlates with their preference for active decision-making and intuition-driven actions (Kehoe & Ludlow, 1988).

Extroverts generally experience more positive emotions and report higher levels of subjective well-being (Lucas & Diener, 2001). They are inclined toward optimism and actively seek new experiences. Introverts, on the other hand, may be more susceptible to negative stimuli and prone to anxiety and depression (Larsen & Ketelaar, 1991).

However, introverts often demonstrate high levels of self-awareness and self-analysis.

Extroverts are more active in social interactions, prefer larger groups, and frequently initiate communication (Ashton, Lee, & Paunonen, 2002). Introverts favor smaller groups or solitude, opting for deep and meaningful conversations over superficial exchanges. Ambiverts, with their behavioral flexibility, can adapt to various social situations, exhibiting either introverted or extroverted traits depending on the context (Revelle & Wilt, 2010).

Eysenck (1967) suggested that differences between introverts and extroverts are related to activation levels of the brain's reticular formation. Introverts have a higher baseline level of activation, making them more sensitive to external stimuli and thus preferring less stimulating environments. Extroverts, with a lower level of baseline activation, seek external stimuli to increase their arousal levels.

Research associates extraversion with the activity of the dopaminergic system (Depue & Collins, 1999). Extroverts exhibit higher sensitivity to rewards, which drives their tendency to seek new experiences and engage in social interactions. Introverts, in contrast, may have lower activity in this system, explaining their preference for calm and solitary activities.

Ambiversion describes individuals who display both introverted and extroverted tendencies depending on the situation (Revelle & Wilt, 2010). Ambiverts demonstrate a high degree of behavioral flexibility, enabling them to adapt to diverse social contexts. They may be sociable and energetic in some situations and calm and reflective in others.

Grant (2013) demonstrated that ambiverts achieve greater success in professions requiring a balance between communication and analytical skills, such as sales and management. Studies also indicate that ambiverts are better equipped to handle changing workplace demands and exhibit greater adaptability compared to strongly introverted or extroverted individuals.

Extraversion is associated with success in professions requiring intensive communication, leadership, and management (Barrick & Mount, 1991). Examples of such professions include management, sales, marketing, and education. Introverts, on the other hand, often excel in fields requiring focus, analytical skills, and independent work, such as scientific research, IT, and the arts. Ambiverts, owing to their flexibility, can successfully work in diverse professional domains, adapting to varying demands and roles (Zelenski, Sobocko, & Whelan, 2014).

Differences in levels of introversion and extraversion are linked to mental health. Introverts may be more susceptible to depression and anxiety, whereas extroverts tend to experience positive emotions and high levels of life satisfaction (Jylhä & Isometsä, 2006). However, excessive extraversion can lead to impulsivity and a propensity for risky behaviors, which also negatively affect mental health.

In the educational context, introverts and extroverts may prefer different learning methods. Introverts may better absorb material through self-study and reflection, while extroverts thrive in group discussions and active participation in the learning process (Zhang, 2008). Understanding these differences enables educators to develop more effective teaching strategies that cater to individual student preferences.

Cultural differences play a significant role in the expression of introversion and extraversion. In collectivist cultures, introversion may be more prevalent and socially acceptable, while in individualistic cultures, extraversion and social activity are often valued (Triandis & Suh, 2002).

These differences should be considered when conducting cross-cultural research and designing intercultural programs.

Genetic studies confirm the heritable nature of extraversion and introversion. Twin studies reveal a high degree of heritability for these traits (Plomin, DeFries, Knopik, & Neiderhiser, 2016). Contemporary genomic research links extraversion and introversion to specific genetic markers, providing deeper insights into the biological foundation of these personality traits.

The advent of digital technologies and social media has created new opportunities for introverts to engage in social interactions in a comfortable environment (Amichai-Hamburger, Wainapel, & Fox, 2002). Extroverts use these platforms to expand their social networks and maintain active social connections. Ambiverts, thanks to their adaptability, effectively utilize digital tools for various purposes, adjusting to the context. Ambiversion is recognized as an important aspect of personality flexibility, enabling effective adaptation to changing conditions and demands.

Research indicates that ambiverts possess a high capacity for emotional regulation and social adaptability, making them valuable employees in dynamic work environments (Grant, 2013).

Understanding the levels of introversion, extraversion, and ambiversion is crucial for career guidance and professional development. Companies employ personality tests to select employees based on the alignment of personality traits with job requirements (Barrick & Mount, 1991). Individuals can use this understanding to choose professions that best suit their personality characteristics and preferences.

Extroverts often thrive in roles demanding communication, leadership, and management, while introverts succeed in areas requiring focus, analysis, and independent work, such as science, technology, and the arts.

In summary, introversion, extraversion, and ambiversion are fundamental personality traits that significantly impact various aspects of life, including professional activities, social interactions, and mental health. Understanding these traits helps individuals recognize their strengths and weaknesses and adapt to different life situations.

Modern research continues to expand our understanding of these complex psychological constructs, integrating biological, cultural, and social factors, which opens new opportunities for practical applications in various fields.

Particularly in the context of scientific processing and researchers' activities, psychological factors and archetypes are critical for maximizing scientific output, achieving valuable results, and advancing scientific careers.

3. **Methodology**

To construct the model of scientists' introversion, multiple linear regression was utilized, enabling the determination of each parameter's contribution to the overall assessment of introversion. The equation includes parameters such as the proportion of solo publications, conference participation, job rank, type of organization, encyclopedic scope of research, depth of research interests, average research duration, citation frequency, publication rate, use of external funding sources, interdisciplinary collaboration, and social media activity. Each parameter contributes to the introversion assessment, weighted by coefficients determined from empirical data.

A comprehensive approach based on Bayesian analysis principles was also employed to evaluate the degree of introversion among scientists. Probabilistic models were used for quantitative analysis, allowing for the assessment of introversion levels based on observed characteristics. Each characteristic was interpreted as a marker of a specific aspect of introversion or extraversion.

Bayesian analysis was selected as the primary methodology due to its ability to integrate prior knowledge and new data for refining estimates. Prior assumptions about the distribution of introversion were based on existing research in psychology and professional behavior. Relationships between characteristics (e.g., frequency of interactions and preference for individual work) were modeled as conditional probabilities. This approach accounted for complex dependencies and enabled a more accurate estimation of introversion levels.

This method facilitated a comprehensive and objective analysis of introversion levels, drawing on both observed data and theoretical concepts.

## 4. Results

The Eysenck Personality Inventory (Eysenck, 1947) is a well-known and relevant test for determining psychological archetypes. However, for this study, we are interested in systems that assess archetypes in a non-test-based mode, focusing on professional interpretation of individual behavioral structures.

The nature of scientific processing suggests that the psychological archetypes of researchers as individuals are particularly prominent in this domain of human activity.

Developing a formalism to define or assess a personality's psychological archetype based on professional activities is a compelling area of interest.

### 4.1. *Linear Formalism of Scientists' Introversion*

To mathematically derive the equation for assessing the degree of introversion *I* among scientists, we define and justify each term in the equation based on hypotheses about the characteristics of introverted scientists.

The introversion equation is constructed as a linear combination of multiple variables, each reflecting a specific aspect of behavior or professional activity associated with introversion.

We assume that a scientist's introversion *I* depends on a set of parameters $x_i$, each contributing to the assessment of introversion.

The contribution of each parameter is regulated by a corresponding weight coefficient $w_i$, which reflects the strength of its influence on introversion.

The equation for *I* is thus expressed as:

$$I = \sum_{i=0}^{n} w_i x_i + \xi$$

where:

$x_i$ - parameters associated with introversion (e.g., proportion of solo publications, conference participation, etc.),

$w_i$ - weight coefficients determining the significance of each parameter,

$\xi$ — random error accounting for unpredictable factors not included in the model (since any procedure for determining or assessing psychotypes is not entirely sterile).

The parameters are introduced, defined, and classified in the following table:

*Table 1.* Scientometric Parameters of Introversion Interpreted Through the Professional Activities of Scientists

| N | PARAMETER | HYPOTHESIS | INTERPRETATION | FORMALIZATION |
|---|---|---|---|---|
| 1 | Share of single publications $PS/PT$, where $PS$ is the number of publications with single authorship, $PT$ - total number of publications | Introverts prefer to work independently, which leads to more solo publications | Share of single publications shows what percentage of publications were created without co-authors. The higher this value, the more likely it is that the scientist is prone to introversion | We formalize the term as $\alpha PS/PT$, where $\alpha$ is a weighting coefficient that determines the significance of the share of single publications. |
| 2 | Number of conferences $C$ | Introverts are less likely to participate in conferences because they tend to avoid public speaking and social interactions. | Fewer conferences $C$ indicates decreased social activity associated with introversion | We formalize the term as $-\beta C$, where $\beta$ is the weight, the minus sign indicates an inverse relationship between conference participation and introversion |
| 3 | Job Rating $R$ | Introverts are less likely to hold high-level positions because it requires management and social skills. | The lower the position (the higher the value of the inverse metric 1-R), the more likely it is that the scientist is prone to introversion | Let's formalize it as $\gamma(1-R)$, where $\gamma$ — weighting coefficient |
| 4 | Organization Type $A$ | Introverts prefer to work in academic institutions, where there is less educational activity than in universities. | Parameter A is equal to 1 for academic institutions and 0 for universities. | Let's formalize it as $\delta A$, where $\delta$ is the weight that regulates the contribution of the type of organization |
| 5 | Encyclopedic nature of research $D$ | Introverts are more encyclopedic because they tend | The higher the D (an encyclopedic score from 0 to 1), the more likely it | Let's formalize it as $\epsilon D$, where $\epsilon$ is the encyclopedicity weight |

| | | to study narrow topics in depth. | is that the scientist is an introvert. | |
|---|---|---|---|---|
| 6 | Depth of research interests G | Introverts focus on deep exploration rather than a variety of superficial topics. | A high G depth score (e.g., a score between 0 and 1) indicates a tendency toward in-depth analysis. | Let's formalize it as ζG, where ζ is the weighting coefficient for the depth of interests |
| 7 | Average duration of studies T | Introverts spend more time researching, immersing themselves in a topic. | A higher T value (in months) indicates long-term studies, which is typical for introverts | Let's formalize it as ηT, where η is the weight of the duration of the research |
| 8 | Citation frequency F | Introverts' work may be less cited because of their narrow specialization | A low citation frequency (1-F) indicates that the scholar is focused on narrow topics | We formalize it as θ(1−F), where θ is the weight for the reverse frequency citation |
| 9 | R Publishing Speed | Introverts publish less often because they prefer to work through the content | A lower R (average number of publications per year) signals less publication frequency. | Let's formalize it as −ιR, where ι is the weight, the minus sign indicates feedback |
| 10 | Use of external sources of financing F | Introverts are less likely to apply for external grants and funds | Low frequency of use of external sources (1−F) may indicate the independence of the scientist | Let's formalize it as −κF, where κ is the weight for using grants |
| 11 | Collaboration with colleagues from other disciplines C | Introverts are less likely to participate in interdisciplinary projects | A low C may indicate a one-way focus and limited interaction. | Let's formalize it as −λC, where λ is the weight |
| 12 | Social Media Activity N | Introverts are less active in professional networks | Low activity in networks (1−N) can be a sign of introversion | Let's formalize it as −μN, where μ is the weight |

Considering all the specified parameters, the equation of introversion can be represented as follows:

$$I = \alpha \frac{PS}{PT} - \beta C + \gamma(1 - R) + \delta A + \epsilon D + \zeta G + \eta T +$$
$$+ \theta(1 - F) - \iota R - \kappa F - \lambda C - \mu N + \xi$$

where all the coefficients α,β,γ,δ,ϵ,ζ,η,θ,ι,κ,λ,μ determine the contribution of each parameter to the model and are subject to calibration using empirical data, and ξ is the random error.

This equation provides a comprehensive evaluation of introversion, considering both preferences in professional activity and the characteristics of social interaction of a scholar.

Here, the proportion of solo publications (PS/PT) and the number of conferences (C) remain key because they are directly related to the level of interaction with colleagues and scientific independence. Additional parameters G,T,F,R,C,N add depth to the model, reflecting such aspects as the depth of engagement, preferences in research methods, level of collaboration, and participation in grant programs.

This model will consider both the personal preferences and level of engagement of the scholar, as well as their interaction with colleagues and involvement in academic communities, enabling a more accurate assessment of the degree of introversion.

Thus, the linear regression model encompasses a broader spectrum of characteristics associated with introversion and can become a more accurate tool for assessing this aspect of personality in scientific professionals.

However, this model is relatively straightforward. At first approximation, it assesses the psychological types of scholars. Yet linearity is often mechanistic: psychological behavioral structures of personality and scientists are sometimes nonlinear and spontaneous. Introversion, extraversion, and ambiversion are pronounced archetypes, but they are sometimes unstable and change over time and circumstances.

Therefore, the degree of introversion is, of course, more of a probabilistic function.

### 4.2. *Bayesian Formalism for Introversion in Scientific Professionals*

We cannot be certain whether professional attributions provide sufficient information about personality and scholar types. We can only interpret this certainty. Statistical interdependence suggests that the optimal construct for interpreting personality types should likely rely on Bayesian formalism, which integrates probabilistic modeling, factor interactions, and uncertainties in the context of scholars' social and professional activities.

Let us consider a Bayesian model that assesses the degree of introversion *I* based on several influencing factors, such as:

P: Proportion of solo publications,
C: Participation in conferences,
R: Professional status,
A: Type of organization,
G: Depth of research interests.

Each of these factors is a random variable with uncertainty, which can vary depending on the context. The Bayesian approach enables the integration of all these factors and accounts for their interactions, creating a probabilistic model for the degree of introversion.

The basic formula of Bayes (Bayes, T., 1763) is represented as follows:

$$P(I \mid D) = \frac{P(D \mid I)P(I)}{P(D}$$

where:
P(I|D): Posterior probability of the degree of introversion I given the data D,
P(D|I): Likelihood, the probability of observing the data D if the degree of introversion is III,
P(I): Prior probability of introversion,
P(D): Normalization constant ensuring the posterior probability is normalized.

Let the data D={PS,C,R,A,G} be a set of factors influencing introversion. For each of these factors, we assume they follow normal distributions dependent on the degree of introversion I. We use a normal distribution to model each factor with parameters dependent on I:

$$P(PS \mid I) = \frac{1}{2\pi\sigma^2(PS)} exp(-\frac{(PS - \mu(PS)^2}{2\sigma^2(PS)}$$

where:
P(PS|I): Posterior probability of the degree of introversion given the proportion of solo-authored publications,
μ(PS): Mean of the proportion of solo-authored publications, which depends on I,
σ2(PS): Variance of the proportion of solo-authored publications (assumed to have a fixed value).

Similarly, for the other factors C (conference participation), R (professional status), A (type of organization), and G (depth of research interests), we get:

$$P(C \mid I) = \frac{1}{2\pi\sigma^2(C)} exp(-\frac{(C-\mu(C))^2}{2\sigma^2(C)}$$

$$P(R \mid I) = \frac{1}{2\pi\sigma^2(R)} exp(-\frac{(R-\mu(R))^2}{2\sigma^2(R)}$$

$$P(A \mid I) = \frac{1}{2\pi\sigma^2(A)} exp(-\frac{(A-\mu(A))^2}{2\sigma^2(PS)}$$

$$P(G \mid I) = \frac{1}{2\pi\sigma^2(G)} exp(-\frac{(G-\mu(G))^2}{2\sigma^2(G)}$$

For each parameter PS, C, R, A, and G, the mean µ will be a linear function of the degree of introversion I, reflecting the relationship between the factors and the degree of introversion. For example, the proportion of solo publications may be directly proportional to the degree of introversion (if no other specific reasons drive solo authorship):

$$\mu = \alpha I + \beta$$

where α and β are parameters that determine the sensitivity of the proportion of solo publications to the degree of introversion. The same assumption can be made analogously for other factors.

Now, suppose the prior distribution of the degree of introversion P(I) is as follows:
1. Uniform distribution: P(I)=1 for I∈[0,1],
2. Normal distribution: If we assume that the degree of introversion in the population of scholars is normally distributed, the prior probability can be represented as:

$$P(I) = \frac{1}{2\pi\sigma^2(I)} exp(-\frac{(I-\mu(I))^2}{2\sigma^2(I)}$$

where μ(I): Mean degree of introversion in the population of scholars, and σ2(I): Variance.

Now we can express the overall posterior probability of the degree of introversion I, given all data and prior assumptions:

$$P(I \mid D) \propto P(I) \cdot P(PS \mid I) \cdot P(C \mid I) \cdot P(R \mid I) \cdot P(A \mid I) \cdot P(G \mid I)$$

Substituting the normal distributions for each factor, we get:

$$P(I \mid D) = \propto \frac{1}{2\pi\sigma^2(I)} exp(-\frac{(I - \mu(I))^2}{2\sigma^2(I)} \cdot \prod_{k \in \{PS,C,R,A,G\}} exp(-\frac{(k - \mu(I))^2}{2\sigma^2(I)}$$

where $k \in \{PS,C,R,A,G\}$.

Using numerical methods for integrating multidimensional distributions (e.g., Monte Carlo methods or gradient descent), we can compute the optimal value of I that maximizes the posterior probability. This value will represent the degree of introversion for a specific scholar, taking all factors into account.

Thus, the proposed Bayesian formalism allows consideration of multiple factors influencing the degree of introversion and their probabilistic distributions. It enables flexible modeling of introversion, considering uncertainties and interactions between factors. The Bayesian approach provides probabilistic estimates, which is a significant advantage when analyzing complex, multifaceted characteristics of scientists as individuals and individuals as scientists.

### 4.3. Simplified Posterior Probability in an Applied Format

With some assumptions, we can simplify the overall posterior probability and present it in an applied and intuitively understandable expression.

Assuming certain approximations, we can represent the posterior probability as a quadratic function of the degree of introversion:

$$P(I \mid D) \propto P(I) \cdot exp(-(A_1 I^2 + A_2 I + A_3))$$

where

$$A_1 = \sum_k \frac{w^2}{2\sigma^2}$$

- coefficient at $I^2$,

$$A^2 = -\sum_k \frac{2w(F-b)}{2\sigma^2}$$

- coefficient at $I$,

$$A_2 = \sum_k \frac{(F-b)^2}{2\sigma^2}$$

- constant,
- F: One of the factors (PS,C,R,A,GP_S, C, R, A, GPS,C,R,A,G),
- $w$ and $b$: Parameters depending on each factor,
- σ: Variance,
- k∈{PS,C,R,A,G}.

To find the maximum of the posterior probability P(I|D), we need to find the value of I that minimizes the quadratic function ($A_1I^2+A_2I+A_3$).

The first derivative is represented as:

$$\frac{d}{dI}(A_1I^2 + A_2I + A_3) = 2A_1I + A_2$$

Setting the derivative to zero:

$$2A_1I + A_2 = 0$$

Obtain for I:

$$I = -\frac{A_2}{2A_1}$$

This is a simple formula for determining the degree of introversion III of a scholar. The quantitative values of internal probabilistic coefficients, parameters, and weights are hidden in the "fog of war," but their interpretation is a matter of numerical modeling and empirical research.

The simplified posterior probability of the degree of introversion III now takes a quadratic form, and optimizing its value reduces to finding the minimum of the quadratic function. This allows for efficient computation of the degree of introversion using standard optimization methods, such as gradient descent or maximum likelihood estimation.

Thus, the Bayesian approach with approximation by normal distributions and linear dependency of factors provides a convenient and intuitively clear method for assessing the degree of introversion in scientists, which can be easily adapted and extended for various tasks.

5. **Discussion**

A linear model enables the assessment of introversion, extroversion, and ambiversion in the context of scientific processing. The key question lies in how we interpret the weighting coefficients embedded within the linear model. Each coefficient in the introversion equation serves as a weight for a specific parameter that reflects the professional and social activities of introverted scientists. These weighting coefficients can be interpreted in a tabular format:

*Table 2.* Interpretation of Scientometric Weighting Coefficients in the Linear Model of Introversion

| No. | COEFFICIENT | EXPLANATION | INTERPRETATION |
|---|---|---|---|
| 1 | α weight of the share of single publications | Reflects the significance of the parameter showing the share of single publications of the scientist. The parameter itself characterizes the scientist's preference for working alone, which is often found in introverts | High value α indicates that the proportion of single publications plays an important role in determining the introversion of a scientist. The higher the proportion of single publications, the more likely it is that the scientist is an introvert. |
| 2 | β weight of conferences | Adjusts the importance of the parameter C, reflecting the number of conferences the scientist has attended. The parameter has an inverse relationship with introversion, since introverts are more likely to avoid public meetings and active social interaction | The higher the value β, the more strongly the model takes conference participation into account as an indicator of extroversion rather than introversion. Low conference participation indicates introversion. |
| 3 | γ job rating weight | Responsible for the significance of the parameter (1−R), where R — job rating. Jobs with higher social skills and leadership requirements are usually filled by extroverts | High value γ indicates that lower positions in the scientific hierarchy are associated with higher levels of introversion. Introverted scientists are generally less likely to hold leadership positions. |
| 4 | δ<br>organization type weight | Regulates the influence of the type of organization A, where the scientist works (1 for an academic institution and 0 for a university). Academic institutions may be more suitable for introverted scientists, as there are fewer teaching responsibilities and public speaking. | If δ is large, this suggests that working in an academic institution is significantly correlated with introversion. |
| 5 | ϵ<br>weight of encyclopedic research | Responsible for the importance of encyclopedic research D. This parameter reflects the narrow or deep nature of the scientist's work in his field. | High value ϵ indicates that encyclopedicity is an important indicator of introversion. The higher the score D, the higher the probability that the scientist is an introvert |
| 6 | ζ<br>weight of depth of research interests | Determines the influence of the depth of research interests G. Introverted scientists tend to focus on deep study of one narrow topic | If ζ is high, it indicates that deep processing of topics is an important indicator of introversion. |
| 7 | η<br>weight of average duration of studies | Adjusts the significance of the parameter T, which shows the average duration of research. Introverts tend to long-term projects | The higher η, the more the model takes into account the length of work as an indicator of introversion, which is typical for introverted scientists who prefer to delve into long-term topics. |
| 8 | θ<br>citation frequency weight | Determines the influence of the parameter ($1-F_c$), where $F_c$ - citation frequency. A low citation rate may indicate a narrow specialization | High value θ suggests that low citation rates may be a sign of introversion, as the scientist's work attracts the attention of a narrow circle of specialists |
| 9 | ι<br>publication speed weight | Responsible for the importance of publication speed $R_p$. Introverts tend to publish less work, focusing on quality | A high ι value indicates that low publication rate is a significant indicator of introversion. |

| 10 | κ<br>weight of use of external sources of financing | regulates the influence of the parameter $F_e$, which shows the frequency of using external financing. Introverts may apply for financing less often | If κ is large, this means that the model considers the frequency of grant use as an important indicator of social activity and, therefore, low activity in this dimension indicates introversion. |
|---|---|---|---|
| 11 | λ<br>the weight of collaboration with colleagues from other disciplines | regulates the value of the parameter $C_d$, which reflects interdisciplinary collaboration. Introverts tend to avoid collaboration outside their discipline | A high λ value indicates that introverts are less likely to work with people from other fields, and this is a significant indicator of introversion. |
| 12 | μ<br>social media activity weight | determines the significance of the parameter $N_s$, which describes the scientist's activity in social networks and professional communities. Introverts usually avoid such activity | The higher the value of μ, the more significant the parameter reflecting the lack of activity in professional networks as an indicator of introversion |

Each coefficient (weight) determines the contribution of its respective parameter to the final introversion score *I*. These coefficients are optimized to minimize the model's error and best reflect the impact of each parameter on the introversion calculation. High coefficient values indicate the importance of the corresponding parameter in characterizing introversion.

In turn, Bayesian analysis is an optimal and powerful tool for determining the level of introversion for several reasons related to its fundamental principles and advantages. Unlike other statistical analysis methods, the Bayesian approach allows for flexible consideration of uncertainties, subjective assumptions, and the updating of knowledge based on new data.

One of the main advantages of Bayesian analysis is its ability to integrate prior knowledge. When working with personality traits such as introversion, we may have subjective assumptions about how different factors influence introversion, based on prior experience or research. The Bayesian approach allows these assumptions to be incorporated through a prior distribution.

For example, if we don't have precise data for a specific scientist, we can use a prior distribution that reflects knowledge of the typical level of introversion in the population of scientists or general observations on the influence of various factors.

Bayesian analysis allows these prior assumptions to be updated as new data becomes available. As a result, it is more flexible than methods that rely only on observed data (such as classical linear regression).

In the context of studying introversion levels, it is important to consider the interdependencies among multiple factors. Bayesian analysis allows these relationships to be expressed using conditional probabilities.

For example, we can model how the proportion of solo publications (PS/PT) depends on the type of organization (A) or how professional status (R) depends on conference participation (C). This enables a more accurate assessment of introversion, taking into account how these factors interact with each other, rather than treating them in isolation.

Unlike other methods, such as linear regression, where dependencies are usually assumed to be fixed and linear, the Bayesian approach allows for more complex and dynamic relationships to be considered.

Bayesian analysis is ideally suited for incremental updates. When new data become available (for example, additional studies on scientists' behavior or new characteristics for already studied scientists), the Bayesian approach allows us to update posterior probabilities without recalculating everything from scratch. This is particularly important in the context of psychology and scientific activity, where new data can significantly impact our understanding.

For example, if future research reveals that conference participation (C) has a greater influence on introversion than previously thought, we can update our prior assumptions and recalculate posterior probabilities for scientists with the new data.

In real-world research, we often face incomplete data. In the case of personality traits such as introversion, precise data on all factors may not always be available. Bayesian analysis provides a natural solution for handling such situations. Even if some data are missing or uncertain, the Bayesian approach can work with prior probabilities and use them to derive posterior distributions.

For instance, if we lack data on professional status (R) for a specific scientist but have abundant information on other factors (such as PS/PT, C, A), Bayesian analysis can still make inferences about introversion based on the available data.

The Bayesian approach provides not only point predictions but also probability distributions for introversion levels. This allows us to obtain more grounded probabilistic estimates, rather than a single number that may be distorted by errors. In real life, when working with subjective and multifaceted traits like introversion, it is important to understand not just the result (e.g., I=0.7), but also the degree of confidence in that result.

This also allows us to account for risk and uncertainty in decision-making, which is important when developing individual recommendations for scientists in their careers or when choosing an appropriate research environment.

Bayesian analysis provides flexibility in model selection. For example, we can use different types of distributions for each factor (normal, log-normal, binomial, etc.), depending on the nature of the data. For instance, if conference participation (C) is a categorical variable (such as "low," "medium," "high"), a binomial or multinomial distribution can be used to model this variable. For numerical variables like the proportion of solo publications (PS/PT), a normal distribution can be applied. This approach allows for more accurate modeling of the real behavior of each factor.

Thus, the Bayesian formalism is a more preferred method for determining introversion levels because it:

*Accounts for prior knowledge and updates it with new data,*
*Considers interdependencies between factors,*
*Provides flexibility in handling incomplete and uncertain data,*
*Gives not only point predictions but also probabilistic estimates, allowing for the assessment of confidence levels,*
*Offers the ability to choose different models for factors.*

These advantages make Bayesian analysis an optimal tool for analyzing subjective and multifaceted characteristics, such as introversion.

However, it is important to note some limitations of the Bayesian approach: One key aspect of Bayesian analysis is the use of a prior distribution P(I), which reflects the initial assumptions of the researcher. If the prior distribution is incorrectly chosen or does not match the real nature of the data, it can substantially distort the results. In the absence of reliable information to select the prior distribution, a uniform distribution is often used, which can be too general and not account for the specifics of the task.

Bayesian methods are most effective when there are enough observations to refine the prior distribution. With small amounts of data, the results may remain uncertain or heavily depend on prior assumptions. When data is sparse, there may be high variability in the posterior distribution, making interpretation difficult.

Bayesian analysis requires calculating the posterior distribution, which often involves complex integrals. These integrals do not always have analytical solutions and may require numerical methods (such as Monte Carlo methods), increasing the computational burden. When working with high-dimensional data or complex models, the computations become more resource-intensive and difficult.

Bayesian analysis also requires constructing a likelihood function, which describes the probability of observing the data given different values of the parameter *I*. If the chosen likelihood model does not correspond to the true data structure, this can lead to incorrect results. For example, using the wrong distributions (e.g., normal instead of exponential) or unsuitable parameters can significantly distort the posterior distribution.

The Bayesian approach gives a probability distribution rather than specific values. This requires additional interpretation (such as calculating means or confidence intervals), which may be unintuitive for non-specialists. The results may also be sensitive to noise in the data, making them less reliable if the data contains many errors or variability. As the number of parameters or dimensions increases, the dimensionality of the task grows, which can lead to problems related to the "curse of dimensionality." This complicates the calculation of posterior distributions and raises the data volume requirements.

Choosing prior distributions and the form of the likelihood function often involves subjective decisions by the researcher, which can influence the final results. If there are no clear theoretical grounds for selecting prior assumptions, the outcome may be subjectively influenced.

If the data does not contain sufficient information to refine the prior assumptions (low informativeness), the posterior distribution may be close to the prior, reducing the practical value of the analysis.

If data on scientific activity (e.g., publication counts) is limited or does not reflect actual behavioral traits, the analysis results may be biased. Missing data on participation in public events could lead to an underestimation of a scientist's extroversion.

The choice of prior distribution (e.g., assuming a uniform distribution for introversion) may not account for the professional or cultural characteristics of the studied group.

Therefore, while Bayesian methods remain a powerful analytical tool, their use requires careful attention to selecting prior assumptions, the likelihood model, and computational methods. To obtain reliable results, it is important to combine Bayesian analysis with additional hypothesis testing and data interpretation techniques.

6. **Conclusion**

This study proposes a linear model for the quantitative assessment of the econometric (scientometric) level of introversion in scientists. The model takes into account various aspects of scientific activity, such as publication patterns, conference participation, job level, and the nature of interactions with colleagues. The development and use of such a model allows scientific institutions to make more informed decisions regarding resource allocation and the creation of environments conducive to the effective work of all employees, regardless of their psychological preferences.

The proposed Bayesian framework also takes into account multiple factors influencing the level of introversion and their probabilistic distribution. This allows for a flexible modeling of introversion, considering uncertainties and interactions between factors. The Bayesian approach provides the ability to obtain probabilistic estimates, which is a significant advantage when analyzing complex and multifaceted personality traits.

The results of the study provide a quantitative assessment of a scientist's introversion level, which can be useful for understanding their professional behavior, work preferences, and interactions with colleagues. Comparing the result with general observations of the scientist's behavior allows for checking the consistency of the model with reality. For example, if an employee actively participates in conferences but the result indicates high introversion, this may signal the need to revise the hypotheses or refine the data.

The derived formulas are interpreted in the context of the professional environment and may not fully correlate with general introversion in other aspects of life (personal, social). Nonetheless, the derived formalism provides a valuable basis for understanding the behavioral traits of scientific workers, but for more accurate interpretation, it is important to consider not only the numerical results but also the context of their application and the limitations of the available data.

Thus, a quantitative approach to assessing introversion among scientists has been developed, based on several hypotheses and parameters of professional and personal activity. The introversion equation, considering factors such as the number of solo publications, conference participation, work type, and the tendency for deep study of narrow topics, has highlighted individual traits characteristic of scientific introversion, extroversion, and ambiversion.

Scientific activity, as a process, is deeply linked to the personality traits and thinking style of the researcher. Individual personality characteristics can have a significant impact on success in various scientific fields. Considering these traits can contribute to more informed career choices and improve the efficiency of scientific work, helping scientists realize their potential in the most suitable environment.

Introverts tend to adopt a deep, analytical approach, prefer solo work, and often exhibit more critical and independent thinking. They feel more comfortable in environments that require minimal social interaction and maximum focus on tasks. Extroverts thrive in dynamic, interactive environments. They tend to have an open mindset and prefer disciplines that involve broad collaboration and social interaction. Ambiverts combine traits of both introverts and extroverts, making their approach to scientific activity versatile. They are comfortable working both independently and in teams, combining deep analysis with a broad perspective.

The proposed linear and Bayesian models adequately describe the introversion of scientific staff, based on a range of parameters of their professional activity. The equations allow for not only the quantitative assessment of introversion but also the identification of aspects where introverts differ from their extroverted colleagues. In particular, parameters such as the proportion of solo publications and participation in conferences were found to be the most significant predictors of introversion. The inclusion of parameters related to social and interdisciplinary interactions also proved to be important. Parameters such as interdisciplinary collaboration and activity in professional networks help to deepen the understanding of the social preferences of introverts and extroverts. This knowledge can be useful for scientific organizations seeking to create favorable conditions for different types of scientists.

The results can be used to adapt the workflow for employees. For example, highly introverted employees can be provided with conditions for focused individual work, while less introverted ones can be given opportunities for increased interaction with colleagues. In team settings, the results can help in role distribution. Introverts may be more effective in analytical or writing tasks, while extroverts may excel in organizational and communication tasks. Scientific institutions can consider the level of introversion when developing training programs, allocating more time for preparation and individual study for introverts.

This aspect is particularly important from the perspective of the ergonomics of scientific processing and the ergonomics of the researcher's scientific activity.

Research on the introversion of scientific workers based on Bayesian analysis has significant potential for application in the fields of econometrics and scientometrics to solve the following tasks:
- Modeling the impact of personality traits on performance,
- In labor economics, it is important to consider the influence of psychological traits, such as introversion, on employee productivity, including scientific workers,
- Models can be applied to assess performance depending on introversion levels, publication frequency, grant activity, and participation in collaborations,

- Optimizing human resource management,
- The use of probabilistic models allows for evaluating how the introversion level of project leaders and participants impacts their success, including project completion on time, grant acquisition, and the number of publications,
- Analyzing the behavior of scientists as economic agents based on their scientometric indicators helps to build more accurate models of scientific activity,
- Models can be used to optimally allocate resources (grants, scholarships) based on personality traits and scientific results,
- Scientific teams can be formed with a balance between introverts and extroverts to increase productivity.

Overall, the research topic on the introversion of scientific workers is significant both for econometrics and scientometrics. The Bayesian approach, applied to the analysis of personality traits, allows, as mentioned earlier, to deepen the understanding of their impact on performance, networking interactions, and career prospects for scientists. This opens new horizons for integrating psychological factors into the quantitative analysis of science and economics.